\documentclass[10pt,conference]{IEEEtran}
\IEEEoverridecommandlockouts
\usepackage{cite}
\usepackage{amsmath,amssymb,amsfonts}
\usepackage{algorithmic}
\usepackage{graphicx}
\usepackage{comment}
\usepackage{textcomp}
\usepackage{xcolor}
\def\BibTeX{{\rm B\kern-.05em{\sc i\kern-.025em b}\kern-.08em
    T\kern-.1667em\lower.7ex\hbox{E}\kern-.125emX}}
\begin{document}
\title{Unburdening onboarding in Software Product Lines}
\author{\IEEEauthorblockN{Raul Medeiros}
\IEEEauthorblockA{University of the Basque Country (UPV/EHU)\\
San Sebasti{\'a}n, Spain\\
raul.medeiros@ehu.eus}
}
\maketitle
\renewcommand\IEEEkeywordsname{Keywords}

\begin{abstract}
The number of studies focusing on onboarding in software organizations has increased significantly during the last years. However, current literature overlooks onboarding in Software Product Lines (SPLs). SPLs have been proven effective in managing the increasing variability of industry software and enabling systematic reuse of a product family. Despite these benefits, SPLs are complex and exhibit particular characteristics that distinguish them from traditional software. Due to these peculiarities, SPLs require a tailor-made onboarding process. Assistance tools might help. In this dissertation, we propose assistance tools (i.e., tools built on top of the software project that help  learners  understand and develop  knowledge) as a means for helping newcomers during onboarding in SPLs.
\end{abstract}
\section{Introduction}\label{intro}
Today's world is highly interconnected. In such a situation, professionals often move from one company to another, especially inside the software industry\cite{SharmaS20}. High mobility implies new hired developers, who must be introduced into the company's culture, processes, technologies, etc., by means of the fastest and most effective possible approach. In the literature, this introduction process is known as \textit{onboarding}\cite{Steinmacher15,Balali18,RastogiT0NC17,bauer11}. Yet, incorporating into a new software development team is not a trivial task, and it has been studied that newcomers (i.e., people incorporating to the company) face several barriers during onboarding\cite{Balali18}, namely:

\begin{itemize}
    \item \textit{technical barriers}, newcomers often encounter themselves without knowing how to set up their development environment or lack the skills to manage the programming languages used in the organization. This problem is accompanied by the lack of prior knowledge of the domain where development takes place. 
    \item \textit{process barriers},  software projects usually lack a formal procedure for introducing newcomers.
    \item \textit{interpersonal barriers}, which refer to the absence of communication and socializing skills in the newcomer. These barriers arise more fragrantly when newcomers are incorporated into a diverse team, where different people with different interpersonal skills gather. 
    \item \textit{personal barriers} include issues related to personality, lack of management skills, or behavior problems. 
\end{itemize}

One of the used approaches to alleviate these barriers consists in considering onboarding as a mentor-newcomer journey (aka mentoring), where the senior developer (i.e., the mentor) assists the newcomer (i.e., the mentee) in her/his incorporation,  both transferring technical skills and knowledge, and providing moral support\cite{SharmaS20}. However, senior developers' time is a valued resource and spending it teaching newcomers reduces significantly their productivity. Therefore, most of the companies expect newcomers to explore and understand the software by themselves\cite{Viviani19}.

Onboarding barriers exacerbate in Software Product Lines (SPL), which are both complex organizationally and software wise. The main goal of SPLs is to support the reuse of a whole family of software products in a systematic way\cite{Clements02}. Specifically, SPLs aim at automating software derivation (i.e., products of the SPL) out of a set of reusable assets (i.e., the SPL platform). Ideally, software derivation is limited to indicating the set of features to be exhibited by the desired products (so-called ``product configuration''). Features are user-visible characteristics of a software system.  In this scenario, products are obtained through a fully ``Configurable Product Family'' where product customization does not exist \cite{Deelstra05}. However, reaching a ``Configurable Product Family'' is estimated to last around ten years\cite{Kolb06}. In this transition, not only traditional onboarding occurs, but also developers are transferred from Application Engineering (AE) to Domain Engineering (DE) (aka \textbf{crossboarding})\cite{Deelstra04}. From now on, we refer to these developers transitioning from AE to DE as \textbf{crosscomers}.

SPLs pose unique challenges to crosscomers, setting the large volume of a SPL aside, crosscomers must adapt themselves to variability and  developing software for reuse. First, feature models gather together multiple concepts that are unknown for the crosscomer, leading to an increase in complexity and making them difficult to comprehend or even maintain\cite{Capilla13}. Moreover, the code should account for reuse. Often, this is achieved by enclosing variants within \#ifdef and \#endif directives, and associated with features. As a result, code comprehensibility is diminished \cite{Ernst02,Melo17}. In summary, the main barriers a crosscomer face when incorporating to the domain engineering team are the sheer  volume,  understanding  variability  and  internalizing  development  for  reuse  \cite{Acher17}.


This thesis focuses on the above-mentioned  barriers. To this end, we resort to assistance tools, which are tools that can help learners understand and develop knowledge (e.g., sensemaking scaffolds\cite{quintana2004scaffolding}). In this context, we pose our research hypothesis as follows: \textbf{\textit{Assistance tools help crosscomers during crossboarding in SPL organizations, simplifying and enhancing the process of incorporating to a domain engineering team}}.

To evaluate the hypothesis, we ask the following research
questions:

\begin{itemize}
    \item \textbf{RQ1:} How can sense-making scaffolds construct domain knowledge on top of crosscomers' background? \textbf{Study-1}
    \item \textbf{RQ2:} What is an appropriate journey for crosscomers to understand variability? \textbf{Study-2}
\end{itemize}
\section{Related Work}
To the best of our knowledge, there is no prior work that addresses onboarding or crossboarding  in SPLs. However, easing and speeding up onboarding has been the subject of distinct approaches:  setting a gamification system \cite{Heimburger20}; tools for assisted coding  \cite{Hibschman19}; visualization of data and conceptual representations \cite{Hibschman19}; recommendation assistants \cite{CubranicM03, Panichella15, MalheirosMTM12,WangS11}; setting a web portal\cite{SteinmacherCTG16}. Specifically, Malheiros et al. \cite{MalheirosMTM12} and Cubranic et al. \cite{CubranicM03} present two tools that support newcomers by recommending the most appropriate source code files for their development tasks. Wang and Sarma \cite{WangS11} focus on bug fixing recommendations, and they provide newcomers a term-based search feature with which newcomers can find bugs on a specific topic. Dominic et al. The \textit{Isopleth} tool \cite{Hibschman19}, the closest approach to this thesis, is a web-based platform that helps newcomers make sense of complex web software projects.


\section{Expected contributions and evaluation plans}

This thesis follows a Design Science Research (DSR) approach. DSR is the scientific study and creation of artifacts as they are developed and used by people with the goal of solving practical problems of general interest\cite{JohannessonP14}. Therefore, the contributions of this thesis have an artifact associated with it. Next, we delve into the proposed studies:

\subsection{Study-1: Domain knowledge construction}
\textbf{\textit{Problem.}} When crosscomers are incorporated into the domain engineering team, they have to deal with both variability and development for reuse. At this point, it is of critical importance that crosscomers understand the concepts and terminology used by the practicioners and how the SPL products are built \cite{Apel13}. However, directly exploring the SPL documentation (e.g. the feature model) or code to acquire domain knowledge is not feasible for crosscomers due to the sheer volume and variability of the SPL\cite{AcherLR18}\cite{Capilla13}. 
\textbf{\textit{Approach.}} In this study, we focus on helping crosscomers understand the set of concepts and terminology of the domain of the SPL platform. To this end, we propose the use of sensemaking scaffolds for presenting such data\cite{quintana2004scaffolding} and specifically concept maps as the realization of such scaffolds. A concept map is a diagram that depicts suggested relationships between concepts\cite{novak2006theory}. 
Concept mapping is reckoned to be a means for meaningful learning insofar as it serves as a kind of scaffold to help to organize knowledge\cite{nesbit2006learning}. Particularly, SPL concept maps gather domain concepts extracted from all available artifacts of the SPL. In our approach, features drive the extraction process. For each feature, its related concepts are extracted from SPL artifacts using NLP techniques such as keyword extraction\cite{AbilhoaC14}. Then, each concept receives a relevance. Finally, an expert selects the most representative concepts of the SPL and creates relationships between them completing the concept map.
\textbf{\textit{Evaluation plan.}} We want to evaluate the effect of concept maps in the acquisition of domain knowledge. This will be carried out with a controlled experiment, with one of our industrial partners, following the experiment design and reporting guidelines proposed by Wohlin et al. \cite{Wohlin12}. To this end, the evaluation will compare the domain knowledge of crosscomers using concept maps with the domain knowledge of those using other kinds of domain representations (e.g. feature models or entity-relationship diagrams), after having performed the given tasks.

\subsection{Study-2: Crossboarding journey}
\textbf{\textit{Problem.}} During crossboarding, crosscomers must move from development with reuse to development for reuse. This means that they have to deal with variability. In SPLs, variability is characterized in terms of  functional abstractions that serve for communicating, reasoning and distinguishing among individual products (i.e. features)\cite{Berger15}. Preprocessors are often the means to implement the features, and has been studied that they hinder program comprehension\cite{Ernst02,Melo17}. Understanding features can be a hard work and it is being reported that onboarding plans that are not structured to newcomers' knowledge, lead to newcomers' frustration\cite{Begel08}. This can also occur in crossboarding and, therefore, crosscomers need to be assigned with features that resemble to their prior knowledge as application engineers. 
\textbf{\textit{Approach.}} In crossboarding, the crosscomer profile can be characterized in terms of the codebase they have been in charge of (i.e. the code they evolved during their time in AE). We refer to this codebase as ``the Background Feature''. Akin to program comprehension theory, we propose to construct crosscomers' journeys (i.e. the order in which they should understand features) based on the similarity between the features and crosscomers' ``Background Feature''\cite{Detienne01}. To define these journeys, we resort to  Recommender Systems (RSs) based on feature similarity. Using Topic Modeling, the RS will create a similarity matrix, which can be represented as a weighted graph. Using this similarity graph, the recommender system will propose the easiest onboarding journey (i.e., a subgraph) based on similarity degrees.
\textbf{\textit{Evaluation plan.}} We plan to evaluate the recommended crossboarding journeys through a case study in one of our industrial partners or in WacLine our academic SPL\footnote{https://onekin.github.io/WacLine/}. The case study will follow Runeson et
al. \cite{Runeson12} guidelines and will try to identify the acceptation and
benefits of the crossboarding journeys in a real SPL crossboarding setting with at least four crosscomers. 


\section*{Acknowledgments}
This work is supported by Spanish Ministry of Science and Innovation (RTI2018-099818-B-I00) and the Ministry of Education (MCIU-AEI TIN2017-90644-REDT (TASOVA)). R. Medeiros enjoys a doctoral grant from the Ministry of
Science and Innovation.
\bibliographystyle{IEEEtran}
\bibliography{20SymposiumICSE}

\end{document}